\documentclass[reprint,prd,aps,showpacs,superscriptaddress]{revtex4-1}

\usepackage[english]{babel}
\usepackage[T1]{fontenc}
\usepackage[latin9]{inputenc}
\usepackage{amsmath, amssymb}
\usepackage{graphicx}


\begin{document}

\title{Isotropic extensions of the vacuum solutions in general relativity}

\author{C. Molina}
\email{cmolina@usp.br}
\affiliation{Escola de Artes, Ci\^{e}ncias e Humanidades, Universidade de S\~{a}o Paulo \\
Av. Arlindo Bettio 1000, CEP 03828-000, S\~{a}o Paulo-SP, Brazil}
\author{Prado Mart\'{\i}n-Moruno}
\email{pra@iff.csic.es}
\affiliation{Instituto de F\'{\i}sica Fundamental, Consejo Superior de Investigaciones Cient\'{\i}ficas\\
Serrano 121, 28006, Madrid, Spain}
\author{Pedro F. Gonz\'{a}lez-D\'{\i}az}
\email{p.gonzalezdiaz@iff.csic.es}
\affiliation{Instituto de F\'{\i}sica Fundamental, Consejo Superior de Investigaciones Cient\'{\i}ficas\\
Serrano 121, 28006, Madrid, Spain}

\begin{abstract}
In this work, we obtain isotropic extensions of the usual
spherically symmetric vacuum geometries in general relativity.
Exact and perturbative solutions are derived. The classes of
geometries obtained include black holes in compact and noncompact
universes, wormholes in the interior region of cosmological
horizons, and anti-de Sitter geometries with excess/deficit solid
angle. The tools developed here are applicable in more general
contexts.
\end{abstract}


\maketitle

\section{Introduction}

Spacetimes described by spherically symmetric solutions of
Einstein's equations are of paramount importance both in
astrophysical applications and theoretical considerations. And
among those, black holes are highlighted. They are relevant
sources of gravitational radiation, offering
possible observational signatures of general relativity
extensions. The current and upcoming gravitational wave
experiments and the possibility of detecting black holes in
accelerators are strong motivations for the investigation of such
models.

In a vacuum, Birkhoff's theorem and its generalizations to
nonasymptotically flat cases uniquely fix the metric as the
Schwarzschild, Schwarzschild--de Sitter or Schwarzschild--anti-de
Sitter geometries, the vacuum solutions of the usual general
relativity with zero, positive or negative values for the
cosmological constant, respectively. Nevertheless, our universe is
not in a vacuum state, even if its dynamics could principally be
driven by an (at least) effective cosmological constant
\cite{Riess,Perlmutter}. Therefore, it is interesting to consider
how the compact solutions in general relativity are modified in a
cosmological scenario. In a different direction, anti-de Sitter
geometries gained interest due to the proposed anti-de Sitter
(AdS)--conformal field theory (CFT) correspondence
\cite{Maldacena,Witten}. This conjecture proposes a duality
between gravity in AdS spaces and CFTs. A better understanding
of AdS backgrounds plays an important role in this program.

In this work we are mainly interested in black holes in a
cosmological environment. Of the two main assumptions of the
cosmological principle, homogeneity is lost when compact objects
are considered. Nevertheless isotropy is still possible, and we
enforce this condition. Within this context, we investigate
spatially isotropic solutions close -- continuously deformable --
to the usual vacuum solutions.

From a more mathematical point of view, isotropy is a condition
that can be implemented in a purely coordinate invariant way, and
as a simple linear constraint in the usual coordinate system
$(t,r,\theta,\phi)$. This geometrical condition is somewhat akin
to the imposition of a constant Ricci scalar, natural in a
Randall--Sundrum brane world context
\cite{Randall-Sundrum,Shiromizu}. Solutions of this constraint can
also be expressed as a linear deformation of the general
relativity vacuum solutions, whose extensions were treated in
\cite{Casadio,Bronnikov2,Bronnikov,Molina1}. 

The present work deals with the problem of extending the usual Einstein's equation solutions in a similar manner. While one practical approach to general relativity is to specify the matter content and from that the spacetime structure, this is not the only possible treatment. For instance, one can specify the spacetime metric based on physical and geometrical considerations, and later follow its implications for the energy-momentum tensor \cite{Casadio,Bronnikov2,Bronnikov,Molina1,Bronnikov:2008,GonzalezDiaz:2011,Morris:1988cz,Morris:1988PRL,Jacobson:2007}. In fact, currently the cosmological scientific community is attempting to adjust the energy density and pressure needed to produce an accelerated FLRW cosmology. We favor here this latter approach. 
 
The structure of this paper is presented in the following. In
Sec. \ref{secII} we develop the basic formalism to be used. In Sec.
\ref{secIII} we apply this formalism deriving linear solutions which
exactly satisfy the isotropy constraint. A large class of
structures appear in the process. In Sec. \ref{secIV}, we go beyond the
linear cases, obtaining more general backgrounds which are
approximately isotropic. Some final comments are made in Sec. \ref{secV}.

\section{General formalism}\label{secII}

In this work, we are interested in spherically symmetric and static
geometries. With these conditions, the metric can be written as
\begin{equation}
ds^{2} = -A(r) \, dt^{2} + \frac{1}{B(r)} \, dr^{2}
+ r^{2} \, \left( d\theta^{2} + \sin^{2}\theta \, d\phi^{2} \right) \,\, .
\label{spherical_geometry}
\end{equation}
In the coordinate system $(t,r,\theta,\phi)$, the stress-energy
tensor has generally the form
\begin{equation}
\left[ T_{\nu}^{\mu} \right] =
\left[\begin{array}{cccc}
-\rho &       &      &     \\
      & p_{r} &      &     \\
      &       & p_{t} &     \\
      &       &      & p_{t}
\end{array}
\right] \,\,.
\label{SET-general}
\end{equation}

We assume spacial isotropy, that is, a linear constraint among $p_{r}$ and $p_{t}$ as
\begin{equation}
p_{r} = p_{t} \equiv p \,\,.
\label{state_equation}
\end{equation}
An eventual non-null cosmological constant is written in the Einstein equations as
\begin{equation}
R_{\nu}^{\mu} - \frac{1}{2} R \, \delta_{\nu}^{\mu} + \Lambda \, \delta_{\nu}^{\mu} = 8\pi T_{\nu}^{\mu} \,\,.
\label{einstein_equations}
\end{equation}
The field equations (\ref{einstein_equations}) imply that the
stress-energy components are related with the functions $A$ and
$B$ as
\begin{equation}
8\pi\rho =  -\Lambda + \frac{1}{r^{2}} - \frac{B}{r^{2}} - \frac{B'}{r} \,\, ,
\label{rho_geral}
\end{equation}
\begin{equation}
8\pi p_{r} =  \Lambda - \frac{1}{r^{2}} + \frac{B}{r^{2}} + \frac{1}{r} \frac{B\, A'}{A} \,\, ,
\label{pr_geral}
\end{equation}
\begin{multline}
8\pi p_{t} =  \Lambda + \frac{1}{2} \frac{1}{r} \frac{A'\, B}{A} - \frac{1}{4} \frac{\left(A'\right)^{2}\, B}{A^{2}} + \frac{1}{2} \frac{1}{r} B' \\
+ \frac{1}{4} \frac{A'\, B'}{A} + \frac{1}{2} \frac{A''\, B}{A} \,\, ,
\label{pt_geral}
\end{multline}
where ``$\,'\,$'' denotes differentiation with respect to $r$. The
Schwarzschild ($\Lambda=0$), Schwarzschild--de Sitter ($\Lambda>0$)
and Schwarzschild--anti-de Sitter ($\Lambda<0$) are the vacuum
solutions of the Einstein equations, trivially satisfying the
condition (\ref{state_equation}) with $p_{t}=p_{r}=\rho=0$ and an
integration constant proportional to a nonvanishing $M$; whereas
for $M=0$, these solutions reduce to the Minkowski, de Sitter, and
anti-de Sitter spacetimes. We are interested in more general
solutions of Eq.~(\ref{state_equation}) which are close to the
Schwarzschild-like solutions, in a natural sense discussed in the
following.

An equation of state in the form of (\ref{state_equation}),
together with the Einstein equations (\ref{einstein_equations}),
imply a functional relation between the functions $A(r)$, $B(r)$
and their derivatives. Using
Eqs. (\ref{pr_geral})-(\ref{pt_geral}), the constraint
(\ref{state_equation}) can be written as
\begin{multline}
r A A' \left(-2\, B + rB'\right) + 2r^{2}AA''B - r^{2}\left(A'\right)^{2}B \\
+ 2A^{2}\left(2 - 2B + rB'\right) = 0
\label{non-pert-exp}
\end{multline}

We now consider the conditions so that our solutions are
continuous deformations of the usual vacuum geometries. Namely, we
assume that the stress-energy tensor $T(r,\delta)$ in the form
(\ref{SET-general}) is a smooth function of a deformation
parameter $\delta$ such that if $\delta=0$ we recover the vacuum
solutions. We assume that the functions $A$ and $B$ are smooth functions of $\delta$, but otherwise unspecified.
Therefore, the stress-energy tensor (\ref{SET-general}) and the
metric components can be written as
\begin{equation}
\left[T_{\nu}^{\mu}\right] = \sum_{n=1} \delta^{n} \, \left[T_{\nu}^{\mu}\right]_{n} \,\, ,
\label{set-expansion}
\end{equation}
\begin{equation}
A = \sum_{n=0} \delta^{n} \, A_{n} \, \, ,
\end{equation}
\begin{equation}
B = \sum_{n=0} \delta^{n} \, B_{n} \,\, .
\end{equation}
The expansions are written so that the constant $\delta$ is dimensionless.

Substituting the form suggested by the stress-energy and metric
elements in the equation of state (\ref{state_equation}), we
obtain for the zero and first order in $\delta$, respectively :
\begin{multline}
r A_{0} A_{0}' \left( -2 \, B_{0} + rB_{0}' \right) + 2 r^{2}A_{0}A_{0}''B_{0} - r^{2} \left( A_{0}' \right)^{2} B_{0} \\
+ 2A_{0}^{2} \left( 2 - 2B_{0} + rB_{0}' \right) = 0
\label{eq-perturb-0}
\end{multline}
and
\begin{widetext}
\begin{multline}
\label{eq-perturb-1}
 A_{0} \left( 2A_{0} + rA_{0}' \right) B_{1}' - \left(
\frac{4A_{0}^{2}}{r} + 2A_{0}A_{0}' + rA_{0}'^{2} - 2rA_{0}A_{0}''
\right)B_{1} \\
=  - 2rA_{0}B_{0} \, A_{1}'' + A_{0}
\left(2B_{0} + \frac{2rB_{0}A_{0}'}{A_{0}} - rB_{0}'\right)A_{1}'
- \left(2rA_0''B_0-2A_0'B_0+rA_0'B_0'+\frac{8A_0}{r}-\frac{8A_0B_0}{r}+4A_0B_0'\right)
A_{1} \,\, .
\end{multline}
\end{widetext}
The requirement that we are dealing with extensions of the vacuum
solutions sets the zero order elements of the expansion as
\begin{equation}
B_{0}(r) = A_{0}(r) = 1 - \frac{2M}{r} - \frac{\Lambda}{3} \, r^{2} \,\, ,
\label{A0}
\end{equation}
which identically solve Eq.~(\ref{eq-perturb-0}). Taking into account the form of $A_{0}$ and $B_{0}$, we treat Eq. (\ref{eq-perturb-1}) writing $A_{1}$ and $B_{1}$ as
\begin{equation}
A_{1}(r) = \left( 1 - \frac{2M}{r} - \frac{\Lambda}{3} \, r^{2}\right) \, a(r) \,\, ,
\end{equation}
\begin{equation}
B_{1}(r) = \left( 1 - \frac{2M}{r} - \frac{\Lambda}{3} \, r^{2}\right) \, b(r) \,\, .
\end{equation}
Substituting in Eq.~(\ref{eq-perturb-1}), we have
\begin{equation}
P_{1}(r) \, b' + 6 b + P_{2}(r) \, a'' + P_{3}(r) \, a' = 0
\label{eq-perturb-1v2}
\end{equation}
with
\begin{eqnarray}
P_{1}(r) & = & -\frac{3r}{2} \left(  2 A_{0} + r A'_{0}\right) = 2
\Lambda r^{3} - 3r + 3M
\label{poly-P1}
\\
P_{2}(r) & = & -3 r^{2} A_{0} = r \left( \Lambda r^{3} - 3r + 6M  \right)
\label{poly-P2}
\\
P_{3}(r) & = & -\frac{3r}{2} \left(  3r A'_{0} - 2 A_{0}   \right)
= 2 \Lambda r^{3} + 3r - 15M \label{poly-P3}
\end{eqnarray}

At this point, the relation (\ref{eq-perturb-1v2}) does not fix a
particular group of solutions which satisfy the isotropy
condition. The complete characterization of $a$ and $b$ is not
possible without additional information. We will develop some
possible classes of solutions in the following sections.

\section{Exact linear solutions}\label{secIII}

Within the presented formalism, we will investigate exact linear
solutions of spherically symmetric and isotropic geometries. They
will provide a rich set of compact structures.

In addition to the already specified requirements --- spherical
symmetry, staticity, isotropy, and a stress-energy tensor
$\left[T_{\nu}^{\mu}\right]$ in the form (\ref{set-expansion}) ---
we will further require that $\left[T_{\nu}^{\mu}\right]$ is
strictly linear in $\delta$, that is, there are no second-order
corrections. Since the exact constraint, Eq.
(\ref{non-pert-exp}), is nonlinear in $A$, a smooth deformation of
$A$ from $A_{0}$ by a small term $\delta\, A_{1}$ would induce
higher-order corrections in $\left[T_{\nu}^{\mu}\right]$. Assuming
strict linearity, and therefore excluding these higher-order
corrections, implies that $A_{1}\equiv0$.

On the other hand, in terms of $B$ the exact relation (\ref{non-pert-exp}) is a
linear first-order differential equation.
Therefore, a smooth deformation of $B$ from $B_{0}$ by an
arbitrarily term $\delta\, B_{1}$ induces strictly linear
corrections in $\left[T_{\nu}^{\mu}\right]$. Indeed, setting
$a(r)\equiv0$ in Eq. (\ref{eq-perturb-1v2}), the general solution
for $b(r)$ can be obtained. We have that
\begin{equation}
b(r) =D\, b_{lin}(r) = D \, \exp \left(  - 6 \int
\frac{dr}{P_{1}(r)} \right)  \,\, , \label{sol_b_linear}
\end{equation}
where $D$ is a dimensionless and positive integration constant. Thus, the functions $A(r)$ and
$B(r)$ can be written as
\begin{equation}
A(r) = A_{0}(r) \,\, ,
\label{sol_A_linear}
\end{equation}
\begin{equation}
B(r) = A_{0}(r) \left[ 1 + C \, b_{lin}(r) \right] \,\, .
\label{sol_B_linear}
\end{equation}
with $C=D \cdot\delta$. The specific choice of $\delta$ is a
matter of convention, since it can be incorporated into $C$. In
the present work, we will be careful to keep the constant $C$
dimensionless.

It is straightforward to verify that the solution
(\ref{sol_A_linear})-(\ref{sol_B_linear}) is exact ($p_{r} \equiv
p_{t}$). Both pressures
will be denoted $p$ as in Eq.~(\ref{state_equation}). Although Eqs.~
(\ref{sol_A_linear}) and (\ref{sol_B_linear}) are valid for any
$C$, we will see that the metric associated with this solution
will describe static and Lorentzian manifolds, for $C$ taking
values in a proper subset of $\mathbb{R}$ only.

One common feature of the obtained spacetimes is that they are characterized by energy densities and pressures which are $r$-dependent, and become constant in the asymptotic region. This qualitative behavior is the expected one for compact solutions immersed in backgrounds which are asymptotically flat, de Sitter or anti-de Sitter. Excluding cosmological constant effects, this phenomenon is also seen when ``hairs'' are present (for example, see \cite{Bizon:1990,Droz:1991,Brady:1997,Molina:2004,Gubser:2008}). Still, the precise form and properties of the obtained geometry are strongly determined by the sign of $\Lambda$. We will treat it then case-by-case in the following subsections.

\subsection{Linear solutions with $\Lambda=0$}

Considering a null cosmological constant, we get for the function $b$
\begin{equation}
b(r) = \frac{(r - M)^{2}}{M^{2}}  \,\, ,
\label{b_Lambda0}
\end{equation}
and for the metric functions
\begin{equation}
A(r) = A_{0} = 1 - \frac{2M}{r} \,\, ,
\label{sol_A_Lambda0}
\end{equation}
\begin{equation}
B(r) = \left( 1 - \frac{2M}{r} \right) \left[ 1 + C \, \left(
\frac{r}{M} - 1 \right)^{2} \right] \,\, . \label{sol_B_Lambda0}
\end{equation}

In order for the geometry described by the functions $A$ and $B$
to have a static region (with $A>0$ and $B>0$), the dimensionless
parameter $C$ must be bounded from below: $-1 < C < \infty$. The
energy density and pressures associated with the metric given by
Eqs.~(\ref{sol_A_Lambda0})-(\ref{sol_B_Lambda0}) are
\begin{equation}\label{rho}
8\pi \, \rho = - C \, \frac{ \left(  r - M \right) \left( 3r - 5M \right)}{(Mr)^{2}} \,\, ,
\end{equation}
and
\begin{equation}\label{p}
8\pi \, p = C \, \frac{ \left( r - M \right)^{2}}{(Mr)^{2}} \,\, ,
\end{equation}
respectively. We also define an equation of state parameter $w=p/\rho$ which is
\begin{equation}\label{w}
w=-\frac{r-M}{3r-5M}.
\end{equation}

As we will show considering particular cases in the following subsections, the energy density and pressure given by Eqs. (\ref{rho}) and (\ref{p}) would generally correspond to the total energy density and pressure which can be obtained when more than one fluid is present. Moreover, as expected from Eq. (\ref{sol_b_linear}), one can
recover the case without deformation with $C=0$. That is, the
functions resulting when $C=0$ is taken in
Eqs.~(\ref{sol_A_Lambda0}) and (\ref{sol_B_Lambda0}) reduce to the
Schwarzshild metric, which corresponds to the vacuum
case.

\subsubsection{Spatially homogeneous and isotropic universes
($M=0$)}\label{A1}

If $M=0$, the metric components are given by
\begin{equation}
A(r) = 1 \,\, ,
\end{equation}
\begin{equation}
B(r) = 1 - K \, r^{2} \,\,\, \textrm{with} \,\,\, -\infty < K < \infty \,\, .
\end{equation}
We have redefined the parameter $C$ into a real parameter $K$ with
dimension of $\textrm{Length}^{-2}$. The metric describes a
\emph{homogeneous and isotropic background} with spacial curvature
$K$.

It can be noted that if we consider $K>0$, we obtain a closed
static model that is the cosmological model studied by Einstein.
In fact, the energy density and pressure given by Eqs.~(\ref{rho})
and (\ref{p}) can be seen as the total energy density and pressure
of a universe filled with two fluids, usual matter with
$8\pi\rho_m=2\,K$ and a cosmological constant with
$8\pi\rho_{\Lambda}=K$, which cancels the gravitational collapse due to the
matter component. Therefore, although we are considering general
relativity without cosmological constant, i. e. $\Lambda=0$ in
Eq.~(\ref{einstein_equations}), we recover the Einstein universe as
a deformation of the vacuum case with $M=0$, the
cosmological constant appears as a fluid which composes part of the
universal content originated by the deformation.

An argument similar to that mentioned above could be applied to
the case $K<0$, which corresponds to an infinite static universe.
Nevertheless, in this case the fluid originated by the deformation
could be decomposed into a negative cosmological constant and matter with $\rho_s<0$ and $w_s=0$. On the other hand, for $K=0$ we
recover the case where no deformation is considered, the Minkowski spacetime.

As we will see, the geometries presented here correspond to the
asymptotic limit of the solutions with $M \ne 0$.

\subsubsection{Black holes in an isotropic and noncompact universe ($M>0$ and
$C>0$)}\label{A2}

For positive values of $C$, the only zeros of the functions $A$
and $B$ are given by $r_{+}=2M$. Moreover, these functions are
analytic and positive-definite for $r_{+}<r<\infty$. Therefore the
coordinate system $\left(t,r,\theta,\phi\right)$ is well defined
in the region $r_{+}<r<\infty$.

The analytic continuation beyond $r_{+}$ is possible with the
standard techniques, and the surface $r=r_{+}$ is a Killing
horizon. It is a simply connected surface for the most simple choice of topology. The interior region have
a curvature singularity for $r\rightarrow0$. On the other hand, in
the limit of very large $r$, or equivalently taking the limit of
zero mass, we obtain an static universe with a negative spacial
curvature. Therefore, the resulting geometry is a \emph{black hole
in a noncompact universe}.

It must be emphasized that we have a static black hole in a
nonempty environment. One possible interpretation of this
geometry can be found noting that the equation of state
parameter, given by Eq.~(\ref{w}), is equal to minus one for
$r=r_+$, and that a black hole in an asymptotically de Sitter
scenario would not accrete a fluid with such a value of the
equation of state parameter (behaving as a cosmological constant)
on its horizon \cite{MartinMoruno:2008vy}. Therefore, when the test-fluid approach used to study the
accretion process \cite{Babichev:2004yx} (and in Ref.
\cite{MartinMoruno:2008vy}) is broken, the most important
characteristics of the fluid are those in the vicinity of the
black hole in any scenario, that is $p+\rho$ on the horizon and
not at infinity. In such a case, a static configuration can be
reached because a black hole would not accrete a cosmological
constant, or any fluid with $w(r_+)=-1$. Nevertheless, it must be pointed out that this would be a
sufficient but not a necessary nonaccretion condition.

\subsubsection{Black holes in a compact universe ($M>0$ and $-1 < C <
0$)}\label{A3}

For negative $C$, that is $-1 < C < 0$, the geometry is more
elaborate. The function $B$ has a second zero $r_{max}$ (besides
$r_{+}=2M$) given by
\begin{equation}
r_{max} = M \left( 1 + \frac{1}{\sqrt{|C|}} \right) \,\, ,
\end{equation}
such that (i) $r_{+}<r_{max}$; (ii) $A(r_{max})>0$; (iii) $A(r)>0$
and $B(r)>0$ for $r_{+}<r<r_{max}$; (iv) $A(r)$ and $B(r)$ are
analytic for $r_{+} \le r \le r_{max}$. As a side remark, we point
out that $r_{max}$ can be arbitrarily large, tending to infinity
as $C$ tends to zero (the Schwarzschild case), or arbitrarily
close to $r_+$ for $|C|\rightarrow 1$ (an extremal
geometry).

The surface $r=r_{+}$ is a Killing and outer trapping horizon, and
the surface $r=r_{max}$ is an inner trapping horizon. It can be
seen that at $r=r_{max}$ the ``flaring-out condition'' which
characterizes wormholes \cite{Morris:1988cz,Morris:1988PRL}, implying the
existence of an outer trapping horizon \cite{Hayward,Prado}, is
replaced by a ``flaring-in condition'' for this inner horizon
case. In order to understand the behavior of the geometry close to
$r_{max}$, we can consider the proper length $\ell$, which is
\begin{multline}
\label{ell}
\ell(r) = \pm 2 M \left|C\right|^{-1/2} \eta \\
\times \left[
\frac{2\left|C\right|^{-1/2}}{\left|C\right|^{-1/2} - 1}
\Pi\left(\mu(r),-\eta,\eta\right)
-\text{F}\left(\mu(r),\eta\right) \right]
\end{multline}
with
\begin{equation}
\eta = \frac{ \left|C\right|^{-1/2} - 1}{\left|C\right|^{-1/2} + 1}
\,\, ,
\end{equation}
\begin{equation}
\mu(r) =
\arcsin
\sqrt{\left( \frac{ \left|C\right|^{-1/2} + 1}{\left|C\right|^{-1/2} - 1} \right)
\,\, \left( \frac{\left|C\right|^{-1/2} + 1 - r/M}{\left|C\right|^{-1/2} - 1 + r/M} \right) } ,
\end{equation}
where $F$ and $\Pi$ denote the incomplete elliptic integrals of
first and third kind, following the conventions in Ref.~\cite{Gradshteyn}, and the expression for $\ell(r)$ is well
defined for values of $r\geq r_+$. We have chosen a function
$\ell(r)$ such that $\ell(r_{max})=0$ and the $\pm$ sign in the
r.h.s. of Eq.~(\ref{ell}) analytic continues the geometry to
negative values of the proper length; thus, one can use the chart
$(t,\ell,\theta,\phi)$, with $-\ell_{max} < \ell < +\ell_{max}$
and $\ell_{max}=\ell(r_+)$ taking a finite value.

\begin{center}
\begin{figure}
\includegraphics[width=0.7\columnwidth]{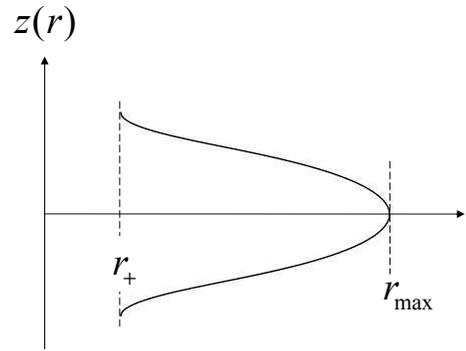}
\caption{Function $z(r)$ which describes the behavior of a section
of the spacetime embedded in an extra dimension. It can be seen
that the geometry flares in at $r=r_{max}$.}
\label{zr}
\end{figure}
\end{center}

\begin{center}
\begin{figure}
\includegraphics[width=0.7\columnwidth]{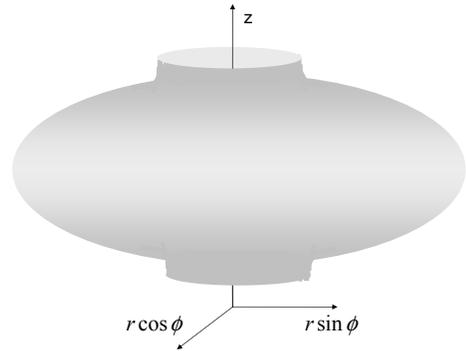}
\caption{Embedded diagram of a section with constant $t$ and
$\theta=\pi/2$. Both charts used to obtain this diagram have
$r_+\leq r\leq r_{max}$, therefore the embedded diagram only
shows regions with $r\geq r_+$.  $r=r_+$ (top and bottom of the
figure) is the limit of validity of this description.} \label{emb}
\end{figure}
\end{center}

Now that we have established the good behavior of the geometry at $\ell=0$, we
can equivalently describe this extension of the space by two
identical charts $(t,r,\theta,\phi)$, both with $r_+\leq r\leq
r_{max}$, which should be matched at $r_{max}$. In order to
visualize this geometry, we can consider that a section of our
spacetime, with constant $t$ and $\theta=\pi/2$, is embedded in an
extra dimension $z$. Therefore, a function $z(r)$ would describe
this section of the spacetime in a higher-dimensional space. The
derivative of such a function would tend to infinity at $r_+$ and
$r_{max}$, and $z(r)$ fulfills the ``flaring-in condition'' in
$r_{max}$ (see Fig.~\ref{zr} ). Thus, this maximal radius would
be a surface of the same kind as that appearing in the equator of
the Einstein static universe, where the light rays are parallel.
In fact, one of the cases appearing in Sec.~\ref{A1} was just the
Einstein universe. It can be seen that also in that case there is
an inner trapping horizon, at $r_*=(K)^{-1/2}$, which is hidden
when one applies the usual change of coordinates,
$r=(K)^{-1/2}\sin\eta$, in order to obtain an extension reflecting
the initial geometry, which corresponds to a spatially spherical
geometry, ${\rm d}s^2=-{\rm d}t^2+K^{-1}{\rm d}\Omega_{(3)}^2$. On
the other hand, in Fig.~\ref{emb} the embedded diagram of this
section of the space is depicted, showing that the geometry
corresponds to a closed universe with two Killing horizons, the
initial one and its reflected image, which plays the role of
upper and lower limits of the figure, respectively. As in the
Einstein model, the region covered by the initial chart is
reflected in an identical region, closing the universe.

\begin{center}
\begin{figure}
\includegraphics[width=0.8\columnwidth]{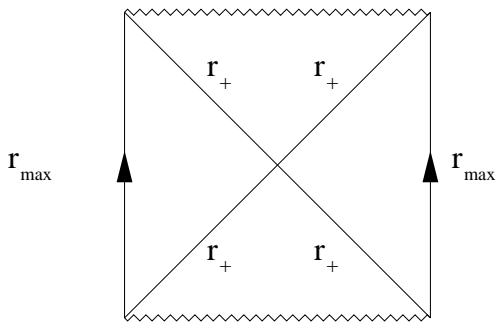}
\caption{Conformal diagram of the maximal extension of the
considered geometry. Lines with arrows are identified. The zigzag
and straight lines denote the interior singularities and the
horizons, respectively.} \label{bh_compact_universe_diagram}
\end{figure}
\end{center}

The Carter--Penrose diagram of the maximal extension of this
geometry can be obtained by considering also values of $r$ such
that $0\leq r\leq r_+$ for both charts, as shown in  
Fig.~\ref{bh_compact_universe_diagram}. As in the Schwarzschild case, for each chart, one obtains a constant radial line at $r=r_+$ which would denote the black (white) hole horizon in the upper (lower) region of the diagram, showing a connection between the corresponding line of each chart.

The difference between this diagram and that of a Schwarzschild spacetime is that, in the
present case, the left-hand and right-hand regions are not ending in the
spatial infinity, but they are identified at a finite maximum
radial radius. Therefore, whereas in the maximal extension of the
Schwarzschild space, one has two asymptotically flat exterior
spaces which present a black hole horizon, in this case, due to the identification at $r_{max}$, there is only one exterior space. Assuming the maximal extension proposed, an observer in the
exterior region can reach the black hole interior region by
crossing two different horizons (each one parametrized by a
different chart of coordinates). Thus, this geometry can be
interpreted as \emph{two black holes in a compact universe} (see
Ref.~\cite{Nandra:2011ug} where a similar interpretation is
considered for dynamical geometries resembling this one). A \emph{one black hole} interpretation is also sensible, based on topological considerations.

Finally, it can be noted that the energy density, pressure and
equation of state parameter, given by Eqs.~(\ref{rho}), (\ref{p})
and (\ref{w}), respectively, take finite and nonvanishing values
outside the black hole, i. e. for $r\geq r_+$. Moreover, $w(r_+)=-1$ as in the
previous studied geometry (Sec.~\ref{A2}), which could be
interpreted as a nonaccretion sufficient condition for an stationary
solution. One could follow an argument similar to that presented
in the case of the Einstein universe studied in Sec.~\ref{A1},
considering that these quantities represent the total energy
density and pressure and the effective equation-of-state
parameter, respectively. In such a case, a possible decomposition
would be to consider one component with $w_m=0$ and another with
$w_\Lambda=-1$, although this second component would not properly
be a cosmological constant because its energy density varies
through the space, $\Lambda=\Lambda(r)$. Thus, one would have
$\rho_m(r_+)=0$ and $\rho_\Lambda(r_+)=|C|/(4 M^2)$; and if
$|C|<1/2$, then $\rho_m$ would start to dominate at $r_*=3M$,
whereas this radius would not be reached for $|C|>1/2$, because
 $r_*>r_{max}$.

\subsection{Linear solutions with $\Lambda>0$}\label{B}

If the cosmological constant is non-null and positive, the
solutions are more complicated. If $0 < \Lambda < 1/9 M^{2}$, then
the functions $A(r)$ and $B(r)$ have two real positive roots $r_{+}$
and $r_{c}$ ($r_{+}<r_{c}$), which are the usual black hole and
cosmological horizons of the Schwarzschild--de Sitter geometry, and
a negative root $r_- $. This is the condition for the
Schwarzschild--de Sitter geometry to be nonextreme, and we will
assume it from now on. For $M>0$, the polynomial $P_{1}(r)$ has
three real zeros $r_{0}, r_{0-}, r_{n}$, with
$r_{n}<0<r_{0-}<r_{0}$, and therefore can be written as
\begin{equation}
P_{1}(r) = 2\Lambda (r - r_{0}) (r - r_{0-}) (r - r_{n}) \,\, .
\end{equation}
The important point is that $r_{0-}<r_{+} < r_{0} < r_{c}$. In
terms of these constants, the function $b$ can be analytically
calculated as
\begin{equation}
b(r) =  C \frac{ \left(r - r_{0-} \right)^{c_{0-}}}{\left(r -
r_{0} \right)^{c_{0}} \left(r + r_{0} + r_{0-} \right)^{c_{n-}}}
\,\, , \label{b_Lambda_positivo}
\end{equation}
where the positive constants $c_{0}$, $c_{0-}$ and $c_{n-}$ are
written in terms of $r_{0}$, $r_{0-}$ and $r_{n-}$ as
\begin{eqnarray}
c_{0} &=& \frac{3/\Lambda}{  (r_{0} - r_{0-})(2r_{0} + r_{0-})}  \,\, , \label{coef_c0} \\
c_{0-} &=&  \frac{3/\Lambda}{(r_{0} - r_{0-})(2r_{0-} + r_{0})}   \,\, ,\\
c_{n-} &=& \frac{3/\Lambda}{(2r_{0} + r_{0-}) (2 r_{0-} + r_{0})}  \,\, .
\end{eqnarray}
The functions $A(r)$ and $B(r)$ are expressed as:
\begin{equation}
A(r) = A_{0} = \frac{\Lambda}{3r} (r_{c} - r)(r - r_{+}) (r -
r_{-})\,\, ,
\end{equation}
\begin{equation}
B(r) = A_{0} \left[ 1 +
C \frac{ \left(r - r_{0-} \right)^{c_{0-}}}{\left(r - r_{0} \right)^{c_{0}} \left(r + r_{0} + r_{0-} \right)^{c_{n-}}}  \right] \,\, .
\label{sol_B_Lambda_positivo}
\end{equation}

Since the coefficient $c_{0}$ in Eqs.~(\ref{coef_c0}) and
(\ref{sol_B_Lambda_positivo}) is positive, the function $B(r)$
diverges near $r=r_{0}$, inside the static region $r_{+} < r <
r_{c}$. This result might suggest that the geometry might not be
well--behaved if $\Lambda>0$. Indeed, we will see in the following
that this is so if $C>0$, when a naked singularity is present. But
for negative values of $C$, we will also see that the geometry is
regular everywhere, describing a wormhole-like spacetime.

\subsubsection{Naked singularities ($\Lambda>0$ and $C>0$)}\label{B1}

If $C>0$, $A$ and $B$ are positive-definite for $r_{+} < r <
r_{c}$. Moreover, $B$ is divergent at $r_{0}$. The geometry is
well defined and static for $r>r_{0}$, but its curvature
invariants are not bounded, as seen by the behavior of the
Kretschmann scalar near $r_{0}$:
\begin{equation}
\lim_{r\rightarrow r_{0}} \left| R_{\alpha\beta\gamma\delta} R^{\alpha\beta\gamma\delta} \right| \rightarrow \infty \,\, .
\end{equation}
Therefore, for this case \emph{a naked curvature singularity} is present at $r\rightarrow r_{0}$.

\subsubsection{Wormholes within a cosmological horizon ($\Lambda>0$ and $C<0$)}\label{B2}

If $C<0$, the function $B(r)$ is not positive-definite between
$r_{+}$ and $r_{c}$: it has a third zero at $r = r_{thr}$. The
relevant points are: (i) $r_{+}<r_{0}<r_{thr}<r_{c}$; (ii)
$A(r)>0$ and $B(r)>0$ for $r_{thr}<r<r_{c}$; (iii) the functions
$A(r)$ and $B(r)$ are analytic for $r_{thr}<r<r_{c}$. Therefore,
the chart $(t,r,\theta,\phi)$ is valid in the region
$r_{thr}<r<r_{c}$. The analytic continuation of this geometry
gives us a wormhole structure, with a throat at $r=r_{thr}$. The
surface $r=r_{c}$ is a Killing horizon in the maximal extension,
and can be interpreted as a cosmological horizon.

The main characteristics of this class of solutions are captured
by the simpler case $M=0$. The metric functions are given by
\begin{equation}
A(r) = 1 - \frac{r^{2}}{r_{c}^{2}} \,\, ,
\label{A_de_sitter_wormhole}
\end{equation}
and
\begin{equation}
B(r) =  \frac{\left(1-r^2/r_c^2\right)\left(r^2/r_{thr}^2-1\right)}{r^2/r_0^2-1},
\label{B_de_sitter_wormhole}
\end{equation}
where $r_{c}^{2}=3/\Lambda$, $r_{0}=r_{c}/\sqrt{2}$,
\begin{equation}
r_{thr}^{2} = \frac{3/\Lambda}{2 (1 - |C|)} \,\, ,
\end{equation}
and $C$ has to take values on the interval $-1/2<C<0$, in order that $r_{thr}<r_c$. In the limiting case $C=0$, which corresponds
to the case without deformation, one has $r_0=r_{thr}$ and,
therefore, $B(r)=A(r)$, consistently recovering the usual de
Sitter metric.

The coordinate system $(t,r,\theta,\phi)$ is valid for $r_{thr} <
r < r_{c}$.  An analytic extension beyond $r=r_{thr}$ can be made
with the proper length $\ell$ as radial function, where
\begin{equation}
\ell(r) = \pm \frac{2|C| r_{c}}{1 - |C|}
\Pi \left( \lambda(r), \frac{1 - 2|C|}{1 - |C|}, \sqrt{1 -2 |C|} \right) \,\, ,
\end{equation}
\begin{equation}
\lambda(r) = \arcsin \sqrt{ \left( \frac{1 - |C|}{1 - 2|C|} \right)
\,\, \left( \frac{r^{2} - r_{thr}^{2}}{r^{2} - r_{0}^{2}} \right)  } \,\, .
\end{equation}
We observe that the geometry described by
Eqs.~(\ref{sol_A_Lambda0})-(\ref{sol_B_Lambda0}) is compact as
expected, that is $-\ell_{max} < \ell < \ell_{max}$ with a finite
value for $\ell_{max}$. The function $\Pi$ is the incomplete
elliptic integral of third kind, following the conventions in
Ref.~\cite{Gradshteyn}. We have chosen $\ell(r)$ such that
$\ell(r_{thr})=0$.

\begin{center}
\begin{figure}
\includegraphics[width=0.7\columnwidth]{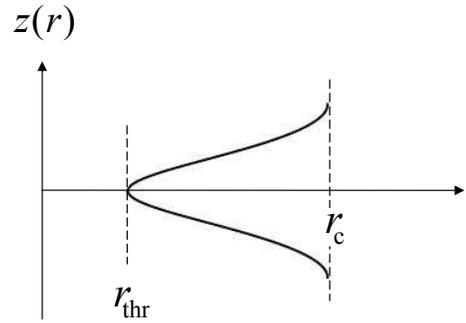}
\caption{Function $z(r)$ which describes the behavior of a section
(with $t=$constant and $\theta=\pi/2$) of the spacetime embedded
in an extra dimension. It can be seen that the geometry flares out
at $r=r_{thr}$ and that the radial derivative of $z(r)$ diverges at both $r=r_{thr}$ and $r=r_{c}$.} \label{zrwh}
\end{figure}
\end{center}

\begin{figure}
\includegraphics*[width=0.7\columnwidth]{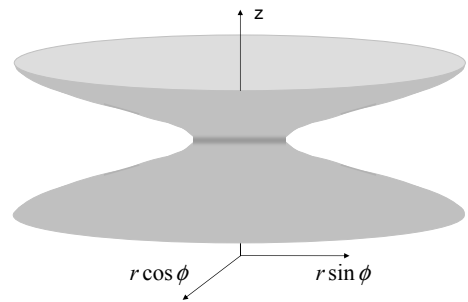}
\caption{Embedded diagram of a section with constant $t$ and
$\theta=\pi/2$. Both charts used to obtain this diagram have
$r_{thr}\leq r\leq r_{c}$.} \label{embwh}
\end{figure}

Following a similar procedure to that presented in Sec.\ref{A3},
once we have checked the good behavior of the geometry at
$\ell=0$, we can describe this extension by two identical charts
$(t,r,\theta,\phi)$, both with $r_{thr} \leq r \leq r_{c}$. The
function $z(r)$ describing the embedding of a section of this
geometry in an additional dimension is shown in Fig.~\ref{zrwh}.
It can be noted that this function fulfills the ``flaring-out condition'' and that whereas the embedding function of an
asymptotically flat wormhole has a radial derivative which
tends to zero for infinitely large values of $r$, in this case,
$z(r)$ is defined only to a finite value of $r$, $r_c$, where its
derivative diverges. In Fig.~\ref{embwh}, the embedded diagram is
depicted; it shows that the surface $r=r_{thr}$ can be considered as a
throat connecting two spatially finite spaces.

\begin{center}
\begin{figure}
\includegraphics[width=0.6\columnwidth]{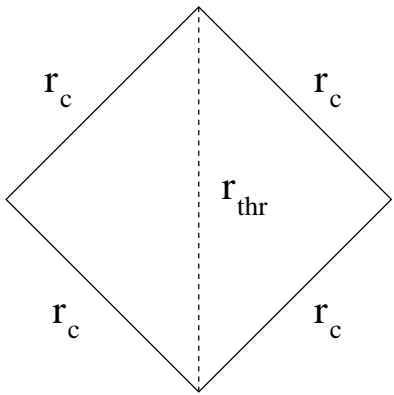}
\caption{Conformal diagram of the wormhole solution inside a
cosmological horizon. Dashed line denotes the wormhole throat.}
\label{de_sitter_diagram}
\end{figure}
\end{center}

In Fig.\ref{de_sitter_diagram}, we present
the Carter--Penrose diagram of the maximal extension of this
geometry. It can be seen that the surface $r=r_{thr}$ ($\ell = 0$)
acts as a wormhole throat and $r=r_{c}$ ($\ell = \ell_{max}$) is
the cosmological horizon. Therefore, this geometry can be
interpreted as a \emph{wormhole-like structure in an
asymptotically de Sitter universe}. That is because the throat connects two universes with a cosmological
horizon at $r_c=\sqrt{3/\Lambda}$ which behave as two
de Sitter universes for large (but still smaller
than $r_c$) values of $r$.

On the other hand, the energy density and pressure can be obtained
taking into account Eqs.~(\ref{A_de_sitter_wormhole}) and
(\ref{B_de_sitter_wormhole}) in Eqs.~(\ref{rho_geral}) and
(\ref{pr_geral}). These are
\begin{equation}\label{rhoB2}
8\pi \, \rho =   - 2 |C | \Lambda \frac{  2  \Lambda ^2 r^4
- 7\Lambda r^2  + 9 }{\left(2 r^2 \Lambda - 3 \right)^2} \,\, ,
\end{equation}
and
\begin{equation}\label{pB2}
8\pi \, p =  2 |C |\Lambda \frac{  \Lambda r^2 -1}{2
r^2 \Lambda -3} \,\, ,
\end{equation}
respectively, leading to the equation-of-state parameter
\begin{equation}
w=-\frac{2\Lambda^2r^4-5\Lambda r^2+3}{2\Lambda^2r^4-7\Lambda r^2+9}\,\, .
\end{equation}
These quantities are finite and nonvanishing
in the interval $r_{thr}\leq r\leq r_c$, and they describe an
matter with $\rho<0$. On the other hand, we have $w(r_c)=-1$
and $-1<w(r_{thr})<0$. That is not in contradiction with the
violation of the null energy condition around and on the throat of
a wormhole needed to maintain such a structure (which is
equivalent to the fulfillment of the flaring-out condition), since
a negative energy density allows $p+\rho<0$ having an equation of
state parameter bigger than minus one. Therefore, the fluid
behaves as dual dark energy \cite{Yurov:2006we} around the
wormhole throat and as a negative cosmological constant on the
cosmological horizon. It should be kept in mind that we are
considering a nonvanishing cosmological constant entering in the
Einstein equations, Eq.~(\ref{einstein_equations}), through the
geometrical part; thus, it is not appearing in the material
content, Eqs.~(\ref{rhoB2}) and (\ref{pB2}), which vanishes for $C=0$.
Nevertheless, that is not in contradiction with a hypothetical
decomposition of the material in two fluids, one of which may be a
positive or negative cosmological constant. In any case, that
second cosmological constant would have a different nature,
because it would be originated by the deformation.

\subsection{Linear solutions with $\Lambda<0$}\label{C}

If $\Lambda<0$ the polynomial $P_{1}(r)$ in Eq.\ref{eq-perturb-1v2} has one real positive root ($r_{0}$) and two complex roots. Therefore, it can be written as
\begin{equation}
P_{1} = - \frac{6}{L^{2}} \left( r - r_{0}\right) \left( r^2 + p r + q \right)
\end{equation}
where we have expressed the (negative) cosmological constant in
terms of the AdS radius $L= \sqrt{-3/\Lambda}$, and $4q - p^{2} =
3r_{0}^{2} + L^{2} > 0$. Using the previous results, we obtain for
the function $b$:
\begin{equation}
b(r) = \frac{\left( r - r_{0}\right)^{c_{0}}  \exp \left[  - \frac{3r_{0} c_{0}}{\sqrt{3 r_{0}^ {2}+ 2 L^{2}}}  \arctan \left( \frac{2r + r_{0}}{\sqrt{3 r_{0}^{2} + 2 L^{2}} } \right)  \right] }{ \left( r^2 + r_{0} r + r_{0}^{2} + \frac{L^{2}}{2} \right)^{c_{0}/2} }  \,\, ,
\label{b_Lambda_negativo}
\end{equation}
where
\begin{equation}
c_{0} = \frac{2L^{2}}{6r_{0}^{2} + L^{2}} \,\, .
\end{equation}
The isotropic linear deformation of the Schwarzschild--anti-de
Sitter geometry is given by
\begin{widetext}
\begin{equation}
A(r) = A_{0} =  1 - \frac{2M}{r} + \frac{r^{2}}{L^{2}} \,\, ,
\label{A_anti_de_sitter_sol}
\end{equation}
\begin{equation}
B(r) =  A_{0}(r) \left\{1 + C
\frac{\left( r - r_{0}\right)^{c_{0}}  \exp \left[  - \frac{3r_{0} c_{0}}{\sqrt{3 r_{0}^ {2}+ 2 L^{2}}}  \arctan \left( \frac{2r + r_{0}}{\sqrt{3 r_{0}^{2} + 2 L^{2}} } \right)  \right] }{ \left( r^2 + r_{0} r + r_{0}^{2} + \frac{L^{2}}{2} \right)^{c_{0}/2} }
  \right\}\,\, .
\label{B_anti_de_sitter_sol}
\end{equation}
\end{widetext}
Because of the complexity of this class of solutions, we initially consider
the case with $M=0$, where
\begin{equation}\label{AC3}
A(r) = 1 + \frac{r^{2}}{L^{2}} \,\, ,
\end{equation}
\begin{equation}\label{BC3}
B(r) = \left( 1 + \frac{r^{2}}{L^{2}}  \right) \left(1 +  C \frac{ r^{2}}{2 r^{2} + L^{2}} \right) \,\, .
\end{equation}
The energy density and pressure associated with this geometry are
\begin{equation}\label{Crho}
8 \pi \rho(r) = -  C \frac{6 r^{4} + 7r^{2} L^{2} + 3 L^{4}}{L^{2} \left( 2 r^{2} + L^{2} \right)^{2}}  \,\, ,
\end{equation}
\begin{equation}\label{Cp}
8 \pi p(r) =  C \frac{ 3 r^{2} + L^{2}}{L^{2} \left( 2 r^{2} + L^{2} \right)}  \,\, .
\end{equation}

\subsubsection{Asymptotically anti-de Sitter space with deficit/excess solid angle ($\Lambda<0$, $M=0$ and $C\geq-2$)}\label{C1}

If $C \ge -2$, then the function $B(r)$ is positive-definite and
the spacetime is noncompact. Its asymptotic limit is not the pure
anti-de Sitter geometry though. Taking the limit $r \rightarrow
\infty$, the line element can be expressed as
\begin{equation}
ds^{2} = - \left[ 1 + \frac{\bar{r}^{2}}{\ell^{2}}  \right] dt^{2}
+ \left[ 1 + \frac{\bar{r}^{2}}{\ell^{2}} \right]^{-1} d\bar{r}^{2}
+ (1 + C/2)\, \bar{r}^{2} d\Omega_{2}
\label{limit_asym_ads}
\end{equation}
after rescaling the radial coordinate as $\bar{r} = (1+C/2)^{-1/2}
r$ and defining $\ell=(1+C/2)^{-1/2} L$, which is the new AdS
radius. For spacetime  described by Eq.~(\ref{limit_asym_ads}),
the solid angle of a sphere of unity radius is $4\pi(1+C/2)$.
Therefore, it presents a solid deficit or excess angle, if the
sign of $C$ is positive or negative, respectively,
\cite{Barriola}. The asymptotic behavior shown in
Eq.~(\ref{limit_asym_ads}) is of a global monopole in an
asymptotic anti-de Sitter spacetime \cite{Li,Bertrand}.

On the other hand, the matter content which leads to this
geometry, Eqs.~(\ref{Crho}) and (\ref{Cp}), strongly depends on
the sign of $C$. If $-2 \le C < 0$, then this material has a positive
energy density and fulfills the null energy condition in the whole
spacetime.

\subsubsection{Compact static universe ($\Lambda<0$, $M=0$ and $C<-2$)}\label{C2}
If $C < -2$, the function $B(r)$ will have a single positive zero $r_{max}$ given by
\begin{equation}
r_{max} = \frac{L}{\sqrt{ |2 + C|} }  \,\, ,
\end{equation}
with $B(r)>0$ for $0 \le r < r_{max}$. Analytic extension can be
made with the proper length as a radial coordinate. In the maximal
extension, the surface $r=r_{max}$ is an inner trapping horizon,
analogous to the already presented case in Sec. \ref{A3}, which
also in this case implies a reflection of the geometry. Therefore,
this spacetime can be interpreted as a compact static universe.

As in the previous case with $-2 \le C < 0$, the matter content of
this spacetime fulfills the null energy condition.

\subsubsection{Geometries with a black hole ($M>0$)}

Considering $M>0$, the new feature is the presence of a
black hole. The functions $A$ and $B$ have a simple positive root
at $r_{+}$, with $0 < r_{0} < r_{+}$. The analytic extension can
be made with the usual techniques and, in the maximal extension,
the surface $r=r_{+}$ is a Killing horizon. It can also be seen
that there is a curvature singularity at $r \rightarrow 0$. On the
other hand, the asymptotic behavior of the spacetime depends, of
course, on the value of $C$.

If $C$ is positive, then the functions $A(r)$ and $B(r)$ are
positive for $r > r_{+}$. The geometry is noncompact with
asymptotic geometry described by the metric
(\ref{limit_asym_ads}). As in the case $M=0$, we observe a solid
deficit angle in the asymptotic limit \cite{Barriola,Li,Bertrand},
whereas for smaller values of $r$ there is a black hole.
Therefore, this geometry can be interpreted as a \emph{black hole
in an asymptotically anti-de Sitter space with deficit/excess
solid angle}.

For negative values of $C$, the function $B(r)$ has another
positive root $r_{max}$, with $r_{+} < r_{max}$ and $B(r)>0$ for
$r_{+} < r <  r_{max}$. As in case discussed for $\Lambda=0$ and
$C<0$, an analytic continuation is possible but the spacetime is
compact. The resulting geometry describes an \emph{anti-de Sitter
black hole in a compact universe}.

Finally, the energy density and pressure of the fluids filling
these spacetimes can be obtained by inserting Eqs.~(\ref{AC3}) and
(\ref{BC3}) in Eqs.~(\ref{rho_geral}) and (\ref{pr_geral}).
Although the obtained functions can not be expressed in a simple
form, it can be seen that they are such that
$w(r_+)=p(r_+)/\rho(r_+)=-1$. Therefore, our hypothesized
nonaccretion condition is again fulfilled.

\section{Beyond the linear solution}\label{secIV}

The developed formalism allows for a great deal of flexibility. We
will focus here on corrections to the linear black hole solutions
discussed in the previous section.

One approach to obtain more general black hole geometries, which
are (approximately) isotropic, is to select an specific choice of
correction function $a(r)$ which satisfies certain physical
criteria. Having $a(r)$ as input, the general solution of
Eq.~(\ref{eq-perturb-1v2}) is given by
\begin{equation}
b(r) = C \, b_{lin}(r)  +  b_{1}(r) \,\, ,
\label{sol_nlin_b}
\end{equation}
\begin{equation}
b_{1}(r) =  b_{lin}(r) \int \frac{P_{2}(r) \, a'' + P_{3}(r) \, a'}{P_{1}(r) b_{lin}(r)}  \, dr \,\, .
\label{b1}
\end{equation}
The polynomials $P_{2}$ and $P_{3}$ are defined in
Eqs.~(\ref{poly-P2}) and (\ref{poly-P3}). The component $b_{lin}$
denotes the linear solution presented in Eqs.~(\ref{b_Lambda0}),
(\ref{b_Lambda_positivo}) and (\ref{b_Lambda_negativo}) for null,
positive and negative cosmological constant, respectively.

With the result in Eq.~(\ref{sol_nlin_b}), the general forms for
$A$ and $B$ are
\begin{equation}
A(r) = A_{0}(r) \left[ 1 + \delta a(r) \right]  + \mathcal{O} \left(\delta^{2}\right) \,\, ,
\label{general_A_nl}
\end{equation}
\begin{equation}
B(r) = A_{0}(r) \left[ 1 + C \, b_{lin}(r) + \delta b_{1}(r) \right] + \mathcal{O} \left(\delta^{2}\right)\,\, .
\label{general_B_nl}
\end{equation}
where $A_{0}$ is presented in Eq.~(\ref{A0}). In Eq.~(\ref{general_B_nl}), the constant $C$ was redefined to absorb a $\delta$ term. We have effectively two deformation parameters: $C$, assuming values in an open set of the real numbers; and $\delta$, such that $|\delta|\ll 1$ to make the perturbative expansion meaningful.

We proceed to the specification of the general form of the perturbation $a$. We will restrict ourselves to the case of null $\Lambda$. Extensions to $\Lambda \ne 0$ are straightforward (but somewhat cumbersome). To ensure that corrections to be introduced do not modify the global proprieties of the solutions already derived in Sec.III, we require that: (i) $a$ should be smooth for $r \ge r_{+}=2M$; (ii) $\lim_{r\rightarrow \infty} a(r) \rightarrow 0$ ; (iii) $a$ should be bounded.

Boundedness of $a$ ensures that the perturbative approach is feasible for small enough values of $\delta$, as will be discussed in the following. If the spacetime spatial section is noncompact, Eq.~(\ref{sol_A_Lambda0}) gives the component $g_{tt}=-A(r)$ of the exact linear metric for $r>2M$.  Since the perturbation $a$ is assumed to be bounded, the more general function $A(r)$ in Eq.~(\ref{general_A_nl}) will remain positive definite for $r>2M$, which is a necessary condition for the staticity of the geometry. In the compact case, the linear solution for $A(r)$ is valid for $2M < r < r_{max}$, but $r_{max}$ can be arbitrarily large. A function $a(r)$ which remains bounded with $r_{max} \rightarrow \infty$ ensures that the more general perturbative $A(r)$ is non-negative. Within these premises, the function $a(r)$ can be written in terms of a set of dimensionless constants $\{a_{1}, a_{2}, a_{3}, \ldots \}$ as an (convergent) inverse power series in the form:
\begin{equation}
a(r) = \sum_{n=1}^{\infty} a_{n} \, \left( \frac{M}{r} \right)^{n} \,\, ,
\label{series_a}
\end{equation}
with $|\delta a_{n}| \ll 1$. This latter condition is compatible with the requirement that the perturbation parameter $\delta$ must be small.

The linearity of the perturbative equation (\ref{eq-perturb-1v2}) allow us to solve it term-by-term. Using the results (\ref{sol_nlin_b})-(\ref{b1}), the solution for $b(r)$ is given by
\begin{equation}
b_{1}(r) = \sum_{n=1}^{\infty} a_{n} \, b_{(n)}(r) \,\, ,
\end{equation}
where the functions $\{b_{(n)}(r)\}$ are
\begin{multline}
b_{(n)}(r) = a_{n} \left[ -(-1)^{n}  \left( n^{2} + 9n + 14 \right) \left( \frac{r-M}{r}  \right)^{2}
\textrm{B} \right. \\
\left. \times \left( - \frac{r}{r-M}; n+2,1-n  \right)
+ (2n+7) \left( \frac{M}{r} \right)^{n}
\right] \,\, ,
\label{func_bn}
\end{multline}
with $ \textrm{B}$ in Eq.~(\ref{func_bn}) being the incomplete beta function, according to the notation in Ref.~\cite{Gradshteyn}. To illustrate the result, the following shows the first few functions in $\{b_{(n)}\}$:
\begin{multline}
b_{(1)}(r) =  a_{1}  \left[
\frac{24 r^2 - 36Mr + 9M^2}{Mr} + 24 \left( \frac{r-M}{M}  \right)^2  \right. \\
\left. \times \ln \left(  \frac{r-M}{r} \right) \right]
\,\, ,
\end{multline}
\begin{multline}
b_{(2)}(r) =  a_{2}  \left[
\frac{ 108r^3 - 162Mr^2 + 36M^2 r + 11M^3 }{Mr^2} \right. \\ 
\left. + 108 \left( \frac{r-M}{M}  \right)^2 \ln \left(  \frac{r-M}{r} \right)
\right]
 \,\, ,
\end{multline}
\begin{multline}
b_{(3)}(r) =      a_{3}  \\
\times \left[ \frac{ 300r^4 - 450Mr^3 + 100M^2 r^2 + 25 M^3 r + 13 M^4 }{Mr^3}
\right. \\
\left. + 300 \left( \frac{r-M}{M}  \right)^2 \ln \left(  \frac{r-M}{r} \right)
\right]
\,\, ,
\end{multline}
\[
\vdots
\]

Although the series in Eq.~(\ref{series_a}) is assumed to be convergent, it is not obvious that the sum in Eq.~(\ref{func_bn}) should also converge. But it indeed does, as can be seen taking the limit of large $n$. We have in this limit
\begin{equation}
\lim_{n \rightarrow \infty} \frac{b_{(n+1)}}{b_{(n)}} = \frac{a_{n+1}}{a_{n}} \frac{M}{r} <  \frac{1}{2} \frac{a_{n+1}}{a_{n}} < 1 \,\, ,
\end{equation}
which is a sufficient condition for convergence.

We finally point that, as required when Eq.~(\ref{series_a}) was proposed, the derived perturbation functions $a$ and $b_{1}$ do not alter the causal and asymptotic characteristics of the linear exact function derived in the previous section.

\section{Conclusions and further comments}\label{secV}

In this work we have shown that the vacuum spherically symmetric
geometries, Minkowski, Schwarzschild, de Sitter, Schwarzschild--de
Sitter, anti-de Sitter and Schwarzschild--anti-de Sitter, can be
isotropically deformed to take into account the existence of some
material content. Even considering linear deformations, in the
sense that the physical quantity $\left[T_{\nu}^{\mu}\right]$ is
strictly linear in $\delta$, we have obtained a zoo of geometries
containing usual or exotic astronomical objects with different
asymptotic behaviors.

In particular, when considering linear deformations of the
Minkowski solution ($\Lambda=0$ and $M=0$), we have obtained
spatially homogeneous and isotropic universes which are spatially
closed (Einstein universe) and open, for negative and positive values
of the deformation parameter ($\delta$, which in Sec. \ref{secIII} is included in $D$, with ${\rm sign}\left(\delta\right)={\rm sign}\left(D\right)$ and, in Sec. \ref{A1}, ${\rm sign}\left(\delta\right)\neq{\rm sign}\left(K\right)$), respectively. It is well known that,
usually in order to consider those models, a cosmological constant
is needed. Nevertheless, in this case the cosmological constant is
not appearing through the Einstein equations, but it is part of
the fluid related to the deformation.

The deformation of a Schwarzschild geometry could lead to a
spacetime where a black hole is in a noncompact universe or two
black holes, the original one and the reflected one, in a compact
universe, depending on the sign of $C$ (which is the same that the sign of $\delta$).
About this second solution, it must be pointed out that we have
been able to obtain a closed structure with two black holes,
because the deformation implies that we are no longer considering a
vacuum background where Birkhoff's theorem holds (as studied
in Ref.~\cite{Uzan:2010nw}).

We have also shown that both de Sitter and Schwarzschild--de Sitter
spacetimes can be smoothly deformed into a geometry which can be
interpreted to describe a wormhole-like structure in an asymptotically de
Sitter universe if $C<0$. Such a wormhole would consistently be
supported by a material content, originated by the deformation,
which violates the null energy condition on and around its throat.

The deformation of the anti-de Sitter geometry leads to a
spacetime which asymptotically behaves as an anti-de Sitter space
with a deficit or excess of solid angle, or a compact static universe, depending on the value of $C$. A black hole should
be considered in those geometries when one is deforming a
Schwarzschild--anti-de Sitter space.

The above mentioned spacetimes which show the presence of black holes are
examples of structures in nonvacuum spacetimes which are in
equilibrium with their environment. We have noted that in all the
studied cases the fluid originated by the deformation is such that
its equation-of-state parameter is equal to minus one in the black
hole Killing horizon. Therefore, we have hypothesized a
nonaccretion condition, in some sense inspired in
Refs.~\cite{Babichev:2004yx,MartinMoruno:2008vy}, which state that
a fluid which behaves as a cosmological constant on the black hole
horizon would not be accreted by it. It must be emphasized that this seems to be a sufficient but not necessary condition in order to have no accretion, at least in principle.

All the geometries obtained consistently
reduce to the vacuum cases when $C\rightarrow 0$. The material
content originated by the deformation seems to
have complicated functional forms in some of the considered backgrounds. Nevertheless, as we have seen in some cases, the energy densities and pressures can be
interpreted as resulting from the sum of two or more fluids with simpler forms.

On the other hand, we have also briefly considered the application
of the deformation formalism relaxing the strict isotropy constraint. As we have shown, in this case additional
specifications are necessary to obtain unique solutions. We
have considered possible choices of the perturbation
based on physical grounds. Nevertheless, it must be pointed out
that this is a powerful formalism which could provide us with new
interesting solutions.

Moreover, although isotropy is the key point in the present work,
extensions of usual solutions in general relativity subjected to
other constraints can be explored within the same approach. For
example, a constraint equation in the form $p_{r}=w_{r}\,\rho$
with a constant value of $w_{r}$ admits as an exact linear
solution (with $\Lambda = 0$)
\begin{equation}
A(r) = 1 - \frac{2M}{r}  \,\, ,
\end{equation}
\begin{equation}
B(r) = 1 - \frac{2M}{r} + \frac{\delta}{r} \left( r - 2M \right)^{-1/w_{r}}
\,\, .
\end{equation}
A rich set of structures can be obtained for the different values
of $w_{r}$ taken. This constraint could be relevant in physical
scenarios.

\begin{acknowledgments}
This work was partially supported by MICINN (Spain) under research
Project No. FIS2008-06332, CNPq (Brazil) and FAPESP (Brazil). C.~Molina also thanks the kind reception by the
group of \emph{Cosmology and Gravitation} at Consejo Superior de
Investigaciones Cient\'{\i}ficas, Madrid.
\end{acknowledgments}

\end{document}